\documentclass{article}
\usepackage{arxiv}

\usepackage[utf8]{inputenc} % allow utf-8 input
\usepackage[T1]{fontenc}    % use 8-bit T1 fonts
\usepackage{hyperref}       % hyperlinks
\usepackage{url}            % simple URL typesetting
\usepackage{booktabs}       % professional-quality tables
\usepackage{amsfonts}       % blackboard math symbols
\usepackage{nicefrac}       % compact symbols for 1/2, etc.
\usepackage{microtype}      % microtypography
\usepackage{lipsum}
\usepackage{graphicx}
\usepackage{url}
\usepackage{import}
\usepackage{tabularx}
\usepackage{booktabs}
\usepackage{amsmath}
\usepackage{amsfonts}
\usepackage{graphicx}
\usepackage{multirow}
\usepackage[table]{xcolor}
\usepackage{rotating}
\usepackage{mathtools}
\usepackage[multiple]{footmisc}
\usepackage{xr}
\usepackage{breakcites}
\usepackage{listings}
\usepackage{color}
\usepackage{hyperref}
\usepackage{authblk}
\usepackage{caption}
\date{}

\title{Identifying most typical and most ideal attribute levels in small populations of expert decision makers: Studying the go/no go decision of disaster relief organizations\\ {\scriptsize PaulIsihara, Chaojun Shi, JonathanW ard, Leo O'Malley, Skyler Laney, Danilo Diedrichs, Gabriel Flores}}

\begin{document}
	\maketitle
	
	\begin{abstract}
	This paper proposes the use of Most Typical (MT) and Most Ideal (MI) levels when an adaptive choice-based conjoint (ACBC) survey can only obtain a small sample size n from a small population size N. This situation arises when expert decision makers are surveyed from among important small populations such as executives of large companies or political leaders, for which the expert decision maker assumption is reasonable. The paper compares respondents’ MT levels obtained using the Build Your Own (BYO) question with MI levels obtained using part-worth utilities. The MI levels are validated using the Potentially All Pairwise RanKings of all possible Alternatives (PAPRIKA) method. It then explores differences in MT/MI levels for two related populations using an application concerning disaster relief. For effective disaster relief coordination, humanitarian organizations must understand each other’s response decisions.  An ACBC survey on the ``Go/No-Go'' decision by 49 faith-based (FBOs) and 12 non faith-based (NFBOs) disaster relief organizations considered four attributes: Funding, Disaster Response Type, Need Assessment, and Community Access. There was disparity between MT/MI Funding levels: 18 of 19 respondents reported MT levels of 50\% or less, but 12 of 19 estimated to have MI levels of at least 75\%. Greatest similarity between FBOs and NFBOs was observed for MI Need Assessment. Greatest disagreement of MI levels determined by part-worths and PAPRIKA was for Need Assessment and Disaster Response Type. To handle zero counts in the sample frequency distributions, we include a mathematical appendix explaining our use of a Bayesian rather than maximum likelihood estimation of MT/MI population frequency distributions.
	 	
	\end{abstract}
	
	\vspace{.5in}
	
	\section{Introduction}
	
		Surveys with a small population size $N$ who are expert decision-makers may arise in a wide variety of important contexts such as a survey of executives of very large companies (Equilar 2019, Russell Reynolds Associates, 2010 ),  cohort of political leaders such as big city mayors (Douglas 2017), or wedding guests' dinner selections (SurveyMonkey n.d.). 	There are a number of practical problems inherent in doing statistical analysis of small populations.   For example, there is the question of wide margins of error when the sample size $n$ is only a fraction of the population size $N$. A 30\% margin of error ($\pm$ 15\% of the true value or proportion) may arise in a business to business  survey sample of $n=10$ companies out of a small known list of $N=100$ companies (Henning 2014). The margin of error reduces to $\pm$ 10\% with a sample size $n=20$, and  $\pm$ 7.5\% with a sample size $n=30$.  This being the case, a recommended approach is to  elicit increased survey response by various methods (phone, e-mail, etc.). 		A different type of small population challenge is encountered in the medical profession (Kirkendall and White 2018).  A small ``hidden'' population' with acute health care needs may involve a much larger, unknown value for $N$ than in a business to business survey population. Members of the population may be largely unknown. Here, small population means relative to the mainstream population, and  the difficulty to gauge, locate, and obtain a sufficiently large sample for statistical inference. In this case,  centralized electronic data collection from multiple sources helps to increase sample sizes (Devers et al. 2014). 	
		
		 In this study we consider two small populations of disaster relief organizations, one faith based, and the other non faith-based. Effectiveness of a disaster response may depend on the quality of collaboration between organizations with a broad diversity of religious and ideological perspectives (Gingerich et al. 2017). For effective coordination of relief, it is important that humanitarian organizations understand the unique traits and characteristics that shape their disaster response decisions (Zakour and Harrell 2003). 
		 
		Faith-based organizations (FBOs) play an increasing role in responding to humanitarian crises, collaborating with larger non-governmental organizations and receiving funding from government programs.  Although many FBO projects have been evaluated for their outcomes, such as the quantity of aid delivered, this survey was part of a larger study to explore how Christian values and perspectives influence the planning and implementation of disaster relief (CCCU n.d.).   Our study provides a data point for the specific question whether and how Christian values influence the decisions made by FBOs when responding to a particular disaster.

		 We designed an ACBC survey (Orme 2014, Ormay and Chrzan 2017, Rao 2014 ) to identify similarities and differences between faith based disaster relief organizations (hereafter abbreviated FBOs) and non-faith based disaster relief organizations (hereafter abbreviated NFBOs) in their ``Go/No-Go decision-making''.  Following the first use of adaptive choice-based conjoint (ACBC) surveys to study priorities in disaster relief logistics (Gralla et al. 2014), our survey focused on four attributes pertinent to the ``Go/No-Go'' decision by disaster relief managers: Available Funding, Disaster Response Type, Assessment of Need, and Access to the Affected Community. We also sought to distinguish Most Typical (MT) attribute levels from Most Ideal (MI) levels for both groups of respondents (FBO and NFBOs).	For example, a large disparity between MT and MI levels was observed in FBO Funding, with all  respondents reporting their MT level was 50\% or less, and more than half estimated to have an MI funding level of at least 75\%.	We used the Build Your Own (BYO) question (Orme and Chrzan 2017, Cunningham, Deal and Chen 2010, Orme and Johnson 2008)  to determine MT levels and binary choice task data (Thurstone 1927, Maydeu-Olivares and B\"{o}ckenholt 2005)  to determine MI levels.

		The two small populations which we surveyed both had a known size $N$. We identified for our FBO population a nearly exhaustive list of $N_1=49$  international FBOs headquartered in the U.S. (one was in Canada). Our NFBO population of size $N_2=12$ consisted of the non faith-based members of National Voluntary Organizations Active in Disaster who engage in international disaster relief. Our sample size $n$ for both populations was small ($n_1=13$ for FBOs and $n_2=6$ for NFBOs),  even though we used phone calls, electronic data collection, and follow-up e-mails in an effort to increase sample sizes.

			The main problem  addressed in this paper is how to validate MI levels obtained using estimated part-worth utilities for a small sample size $n$.  Questionable use of part-worths is exemplified by an ACBC survey with a sample size $n=9$ which reported a hierarchical Bayesian estimated part-worth  choice task hit rate of 53.5\% compared to 52.8\% for part-worths estimated by Monotone Regression  (Lighthouse Studio Help n.d.). The solution we propose is to validate part-worth utilities using a basic  PAPRIKA (Potentially All Pairwise RanKings of all possible Alternatives) method (Hanson and Ombler 2009) whose judgments (choice tasks) are structured as and limited to a single-elimination round of 16 tournament. PAPRIKA is particularly appropriate for small samples whose respondents are all expert decision makers whose choice task data is non self-contradictory.  In our case, 16 out of 19 respondents each had over 15 years experience as disaster relief managers.  The probability that PAPRIKA correctly identifies an expert respondent's MI levels depends on the structure of the survey and choice task tournament, but not on the sample size $n$. Thus, PAPRIKA is an effective validation tool for small samples. In our survey,  PAPRIKA confirmed part-worth estimated similarity between FBOs and NFBOs   for MI Need Assessment, with ``Clear Need'' for the organization's services being the MI level for almost all respondents.  	Major disagreement between PAPRIKA and part-worth utility based MI levels identified specific areas where further investigation of an attribute or level is necessary.  In our case, our analysis revealed two such areas: (i) the most ideal disaster response type attribute and (ii) the MT/MI local partner level of the Community Access attribute for both FBOs and NFBOs.
			
				 Observed differences in MT and MI levels between FBOs and NFBOs suggest possible areas for further research into how Christian values may influence FBO decision-making.  For example, a much larger proportion of FBOs (96\%) than NFBOs (50\%) is estimated to respond to  ``non-headline news'' type disasters, meaning those which are not declared ``Level 3'' by the United Nations' Inter-agency Standing Committee (i.e. ``IASC 3'').  This may reflect an FBO value to avoid the limelight in doing good works as was expressed by one of the FBO disaster relief managers.  Another observed difference between FBOs and NFBOs  was in the PAPRIKA  MI funding level. Whereas 1/3 of the FBO's MI level was 25\% or less, this level was a zero count for the NFBOs.  This agreed with  a preliminary study (Veatch and Reimel 2017) which suggested for FBOs (i) an emphasis on reaching the most vulnerable or remote people, and a de-emphasis of cost or efficiency considerations; and (ii) a desire to respond to disasters according to the needs, even when donors are not interested in a particular disaster.

 The layout of the paper is as follows: Section 2 describes the ACBC survey instrument (2.1), data collection and processing (2.2), and PAPRIKA validation of partworth utilities (2.3); Section 3 summarizes results for both FBOs and NFBOs, including comparison of MT and MI levels, comparison of partworth MI and PAPRIKA MI levels, and estimation of MT level proportions for the entire population (Table 3); Section 4 describes a Monte Carlo simulation to estimate the probability that PAPRIKA correctly determines MI levels; and finally,  Section 5 summarizes the MT/MI estimation method as presented, and how the method might be improved. 	As there are a number of acronyms used in the sequel, we compile them alphabetically in Table 1 for quick reference.

	\begin{table}[t]
	\scriptsize
	\captionsetup{width=\linewidth}
	\caption{\\Abbreviations.\hspace{6in}.}
	\centering
	\label{acr}
	\begin{tabular}{lll}
	Acronym & Unabbreviated & Description\\	\hline
	ACBC & Adaptive Choice-Based Conjoint & survey type used to collect choice-task data (Sec. 2.1)\\
	BYO & Build-Your-Own & question within an ACBC survey (Fig. 2)\\
	FBOs & Faith-Based Disaster Relief Organization & one of two targeted U.S.-based survey populations with size $N_1=49$\\
	IASC & Inter-Agency Standing Committee & United Nations agency determining the scale of a disaster response (Fig. 1, Table 3)\\  
	MI & Most Ideal & respondent's most ideal attribute level (Table 2) \\
	MT & Most Typical & respondent's most typically encountered attribute level (Fig. 2, Table 3)\\
	NFBOs & Non Faith-Based Disaster Relief Organization & one of two targeted U.S.-based survey populations with size $N_2=12$\\
	PAPRIKA & {\bf P}otentially {\bf A}ll {\bf P}airwise {\bf R}an{\bf K}ings of all possible {\bf A}lternative & established method proposed here for small sample part-worth validation (Sec. 2.3)\\
	WMAE & Weighted Mean Absolute Error & error measure used for zero-count estimation (Appendix)
	 
	\end{tabular}
\end{table}

	\section{Methods}
	\subsection{ACBC survey platform}

\begin{figure}[!htpb]
		\centering
		\includegraphics[width=5.75in, height=5.95in]{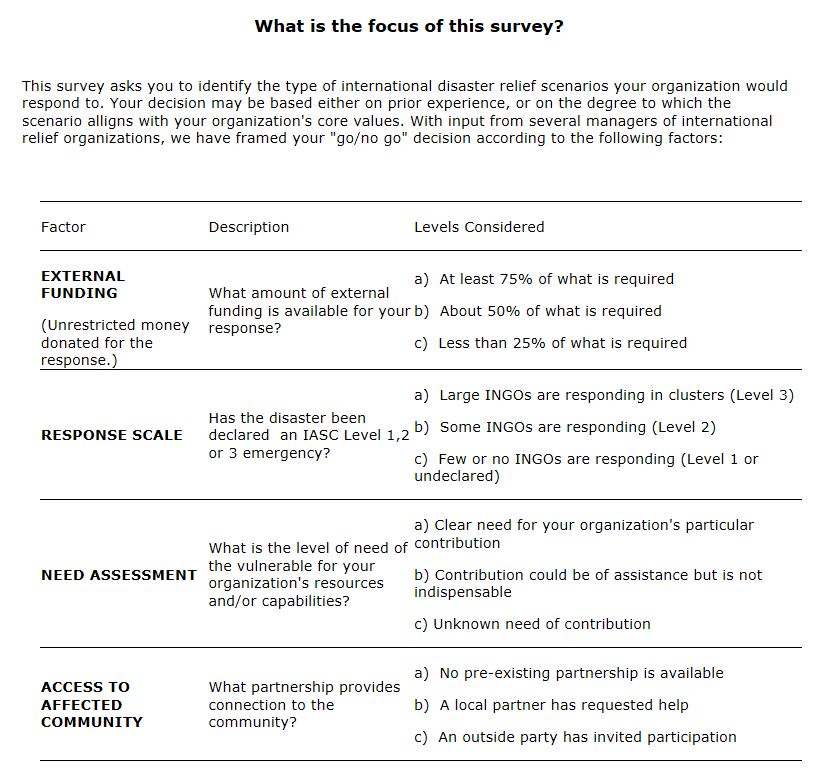}
		\caption{{\small Screen shot of a recent ACBC survey with 4 attributes consisting of 3 levels each.  (INGO=International Non Governmental Organization.)   }}
		\label{Survey}
	\end{figure}

	 After an extensive search and interviewing of international disaster relief organizations with headquarters in the US (we included one with headquarters in Canada), an ACBC survey on the ``Go/No-Go'' decision  was first deployed in Summer 2018 to  a small population  of $N=49$ faith-based disaster relief organizations (FBOs). In Fall 2018, the same survey was administered to an even smaller population of $N=12$ non-faith-based international disaster relief organizations (NFBOs) obtained from a membership list of National Voluntary Organizations Active in Disaster. Our ACBC survey considered 4 attributes (funding, disaster response scale, need assessment, and community access), with each attribute having 3 levels (Figure \ref{Survey}).

\begin{figure}[t]
		\centering
		\includegraphics[width=4.75in, height=3in]{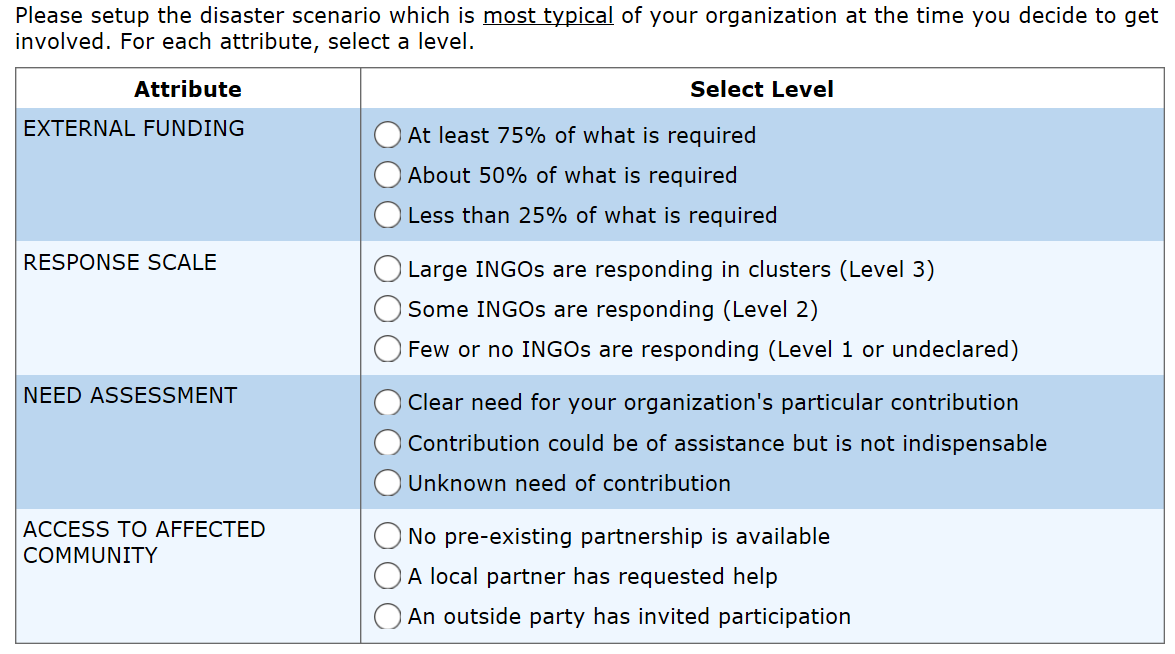}
		\caption{{\small Screen shot of  a non-standard ``Build-Your-Own'' (BYO) question asking respondents to identify the most typical (MT) levels (rather than most ideal (MI) levels) at the time of the Go/No-Go decision. }}
		\label{BYO}
	\end{figure}

	 We employed a non-standard use of the  ``Build Your Own'' (BYO) profile question to determine most typical (rather than most ideal) attribute levels  at the time of the ``Go/No-Go'' decision (Figure \ref{BYO}).  	For each respondent, our online survey platform (Sawtooth's Lighthouse) automatically generated 24 profiles by randomly varying exactly 2 of the respondent's 4 BYO levels.   Including the BYO, these 25 profiles became candidates for the choice task tournament, out of which  16 profiles were randomly selected. (In special cases, the screening stage might reduce the number of profiles included in the tournament as a result of ``must have'' or ``totally unacceptable'' levels.) 
	 
\begin{figure}[b]
		\centering
		\includegraphics[width=4.75in, height=2.in]{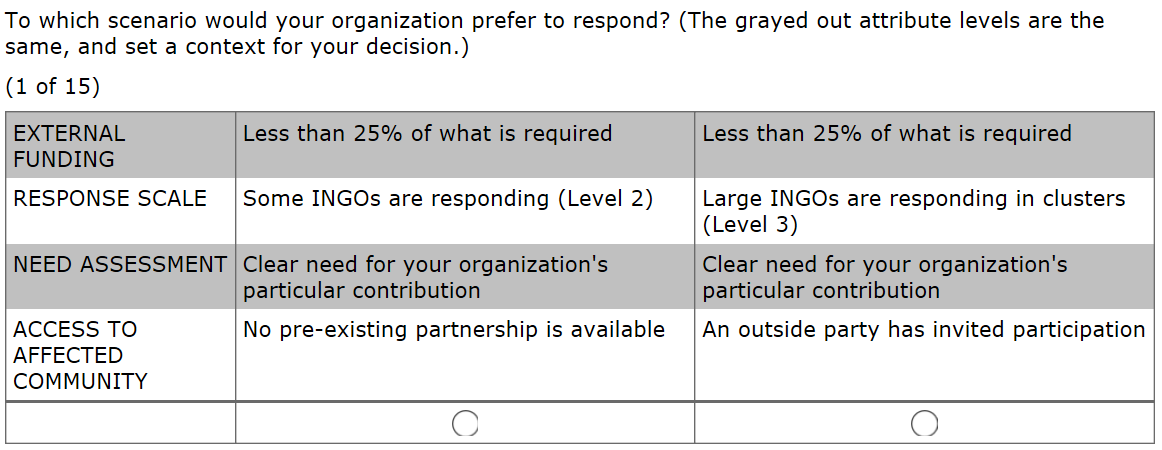}
		\caption{{\small Screen shot of a paired comparison choice-task asking respondents to indicate their preferred profile. Questions were organized as a single elimination tournament beginning with 16 profiles, with selected profiles deemed more ideal.}}
		\label{CT}
	\end{figure}

	 Two different disaster profiles were paired off  in each choice task (Figure \ref{CT})  within a single elimination tournament similar to the FIFA World Cup Round of 16.  Binary choice was deemed easiest for the respondent to determine which among a given set of profiles, varied from the one most typically experienced, is most preferable. The respondent's choices of more ideal profiles advanced to the next round of the tournament. A single elimination tournament structure removes much of the possibility for contradictory responses.  This type of survey structure, combined with the knowledge that 16 of 18 respondents are expert decision makers with over 15 years experience as disaster managers, facilitated use of the PAPRIKA method 	(Hansen and Ombler 2009) to validate estimated dominant partworth utilities obtained by a Hierarchical Bayesian Markov Chain Monte Carlo analysis (Rossi et. al. 2005).

		\subsection{Data collection and Processing}
		
		ACBC surveys  are typically administered to large populations, with a sample size $n$ usually ranging from 300 to 1,200. Small population surveys have been defined formally as follows (Lighthouse Studio Help n.d.). Let $n$ be the number of survey respondents, $t$ the number of choice tasks, $a$  the number of alternatives per choice task, and $c$ the maximum number of levels in any attribute. A small ACBC study is such that $n < N < \frac{c}{at} \cdot 10^3$  (Orme 2014).  This was the case for our parameters $a=2,t=15,c=3$, with $n=13$, and $N=49$ for the FBO population and $n=6$, $N=12$ for the same survey given to NFBOs. 
	
  Our population of $N=49$ faith based organizations (FBOs) spread across the US/Canada and is not accountable to a single administrative body. As such, an online ACBC survey instrument proved to be a useful tool to collect and analyze data (Sawtooth Software Inc. 2017).

  The main goal of data processing was to compile for both FBOs and NFBOs sample frequency distributions of MT and MI levels.  MT levels were simply obtained as responses to the BYO question. MI levels, on the other hand, were obtained as dominant part-worth utilities obtained by a sophisticated hierarchical Bayesian  Markov Chain Monte Carlo  analyzer which is integrated into our ACBC online platform.  Validation of the MI levels became the main methodological question as discussed in the next section.

	\subsection{PAPRIKA Validation of Partworth Utilities}

	Sophisticated Hierarchical Bayesian methods have been integrated into ACBC software to determine individual respondent part-worth utilities from choice task data (Sawtooth Software 2009). Such Bayesian methods share information about all sample respondents to improve estimation of each respondent's part-worth utilities.  As part-worths computed from very small samples may have low hit-rates on small sample choice task data, there is a need to validate the dominant part-worths used to determine MI levels. The Potentially All Pairwise RanKings of all possible Alternatives (PAPRIKA) method with judgments (binary choice-tasks) specified  by an ACBC choice task tournament offers a simple approach to such validation.

	The PAPRIKA method begins with all possible attribute level rankings, and then uses choice task data to pare down the possible rankings to a subset of ``feasible rankings''. For a simple example of this method, consider a survey with two attributes, $A$ and $B$, where $A$ has three levels, namely $A_1$, $A_2$, and $A_3$, and $B$ has two levels, $B_1$ and $B_2$. In this case, as shown in Table \ref{acr}, there are twelve possible distinct attribute level rankings, where levels of each attribute are listed from least to most preferred level. For example, the fifth ranking on the list ($A_2 A_1 A_3 B_1 B_2$) indicates that for attribute $A$, level $A_2$ is lowest ranked, $A_1$ is in the middle, and $A_3$ is highest ranked (most ideal). Similarly, $B_2$ is ranked the most ideal (MI) level for attribute $B$. Given choice task tournament results, we can reduce the number of feasible  attribute level rankings. For example, let us consider the following two tournament results: i) $A_1 B_2$ wins over $A_2 B_2$ and ii) $A_3 B_1$ wins over $A_3 B_2$. Thus, even without any specific knowledge about the utility values, we know that: 
	
	\begin{equation}
	\upsilon(A_1)+ \upsilon(B_2)>\upsilon(A_2)+\upsilon(B_2) \text{ and} 
	\end{equation}
	
	\begin{equation}
	\upsilon(A_3)+ \upsilon(B_1)>\upsilon(A_3)+\upsilon(B_2),
	\end{equation}
	
	where $\upsilon(x)$ denotes the partworth utility of level $x$. This rules out all rankings that do not satisfy these two inequalities. Consequently, we can eliminate as unfeasible 9 out of the original 12 possible attribute level rankings, 	leaving just three feasible rankings, namely 	$A_2 A_1 A_3 B_2 B_1$,  	$A_2 A_3 A_1 B_2 B_1$ and 	$A_3 A_2 A_1 B_2 B_1$. Observe that $A_1$ is the top ranked in two of the three feasible rankings and $B_1$ is top ranked in all three. Thus, without any additional tournament data, we would conclude that $A_1$ and $B_1$ are the respondent's most ideal (MI) levels. 
\begin{table}[t]
	\small
	\captionsetup{width=.71\linewidth}
	\caption{\\All possible orderings of attribute levels, and those eliminated by the choice task data i) $A_1 B_2$ wins over $A_2 B_2$; and ii) $A_3 B_1$ wins over $A_3 B_2$.}
	\centering
	\small
	\vspace{.2in}
	\label{acr}
	\begin{tabular}{|c|c|c|}
		\hline
		Attribute 1 & Attribute 2 & Eliminated (X) \\
		\hline
		\hline
		$A_1 A_2 A_3$ & $B_1 B_2$ & X\\
		\hline
		$A_1 A_2 A_3$ & $B_2 B_1$ & X\\
		\hline
		$A_1 A_3 A_2$ & $B_1 B_2$ & X\\
		\hline
		$A_1 A_3 A_2$ & $B_2 B_1$ & X\\
		\hline
		$A_2 A_1 A_3$ & $B_1 B_2$ & X\\
		\hline
		$A_2 A_1 A_3$ & $B_2 B_1$ &  \\
		\hline
		$A_2 A_3 A_1$ & $B_1 B_2$ & X\\
		\hline
		$A_2 A_3 A_1$ & $B_2 B_1$ &  \\
		\hline
		$A_3 A_1 A_2$ & $B_1 B_2$ & X\\
		\hline
		$A_3 A_1 A_2$ & $B_2 B_1$ & X\\
		\hline
		$A_3 A_2 A_1$ & $B_1 B_2$ & X\\
		\hline
		$A_3 A_2 A_1$ & $B_2 B_1$ &  \\
		\hline
\end{tabular}
\end{table}

PAPRIKA can be applied to surveys with greater numbers of attributes and levels.  For the case of our disaster relief survey survey with 4 attributes each having 3 levels, there are 6 possible orderings of levels in each attribute, and hence a total of $6^4=1296$ possible attribute level rankings.  For example, if we designate the attributes $A,B,C$ and $D$ and the categorical levels using a subscript $1,2$ or $3$, then one possible respondent ranking (from lowest utility to highest utility in each attribute) is $A_1,A_2,A_3,B_1,B_3,B_2,C_3,C_1,C_2,D_2,D_3,D_1$.
Note that, in this case, the respondent's highest ranked levels are $A_3, B_2, C_2$ and $D_1$. (Our discussion focuses exclusively on the highest ranked levels, but the PAPRIKA validation method applies to the middle and lowest ranked levels as well.)

Under the assumption that an expert respondent answers each question consistently, each choice task outcome essentially places a constraint on the respondent's set of possible rankings. For example, suppose the respondent prefers profile $A_1
B_2C_1D_1$ over $A_2B_2C_2D_1$.  Even without any specific knowledge about the utility values, we know that
\begin{equation}
\upsilon(A_1)+ \upsilon(C_1)>\upsilon(A_2)+\upsilon(C_2),
\label{MIinequality}
\end{equation}
where $\upsilon(x)$ denotes the partworth utility of level $x$.
This would then rule out all rankings that do not satisfy
the inequality (\ref{MIinequality}).  For example, any ranking which includes both $A_1A_2A_3$ and $C_1C_2C_3$ would not be feasible. 	The rankings which remain after all the choice task tournament outcome inequalities such as  (\ref{MIinequality}) have been processed, forms the respondent's feasible ranking set. For simplicity, the MI level for each attribute is the level which appears most frequently as the highest ranked level in the respondent's feasible ranking set. The MI level sample frequency distribution  summarizes the number of occurrences that each level is ranked highest by the sample respondents.

	\section{Results}
		Our survey results are summarized in Table \ref{MTMI}.  Section (A) of this table compares MI levels with MT levels. For FBOs, only Community Access showed agreement between MT and 	MI levels, with Access through local partners almost unanimous for both.  All 13 FBOs  most typically had 50\% or less funding in hand  at the time of the Go/No-Go decision, but over 50\% indicated their ideal funding level was at least 75\%. While a large majority FBOs most typically responded to ISAE Level 2 disasters, 30\% indicated responding to the largest scale disasters (ISAE Level 3) as their ideal. 	Whereas FBOs most typically were evenly split between ``clear need'' and ``optional''  Assessed Need for their services, almost all viewed "clear need" as the ideal. In the case of NFBOs, Assessed Need showed the greatest agreement between MT and MI levels, with ``clear need'' almost unanimous for both. Similar to FBOs, almost all NFBOs were most typically funded at 50\% or less, and most ideally would have at least 75\%  in hand at time of the Go/No-Go decision. While half the NFBOs responded to smaller scale disasters, all viewed the largest scale response (IASC 3) as their MI level.  Similarly, while half the NFBOs most typically did not operate through local partners, all viewed local partners as their most ideal Community Access.

					\begin{table}[t]
						\captionsetup{width=.95\linewidth}
						\scriptsize
						\caption{{\small \\(A) Build-your-own (BYO) determined MT level and hierarchical Bayes estimated part-worths based MI level sample frequency distributions. (B) MI sample frequency distributions obtained by hierarchical Bayesian  estimated part-worths and by the PAPRIKA method.  (C) Estimation of MT level proportions for the total population. Zero counts in the sample are estimated by a Bayesian estimation method (see Appendix).} }
						\centering
						\label{MTMI}
						\begin{tabular}{ll||cc|cc}
							{\bf (A)} &&	\multicolumn{2}{c}{FBO} & \multicolumn{2}{c}{NFBO}\\
							Attribute	&Level&BYO MT & part-worth MI	& BYO MT & part-worth MI\\\hline
							& $\ge 75\%$& 0 & 7 & 1 & 5 \\
						Funding&$\sim 50\%$ & 6 & 2& 2 & 0 \\
							&$<25 \%$ & 7 & 4 & 3 & 1 \\\hline
							& IASC 3 & 0 &	4 &	3 &	6 \\
							 Disaster Response Type&IASC 2&11&6 &2 & 0\\
							& IASC 1 or undeclared &2 & 3 & 1 & 0 \\\hline
							& clear & 6 &12&5&6\\
						 Assessed Need& optional&7&1&1&0\\
							& unknown&0&0&0&0\\\hline
							&none&0&0&2&0\\
							 Community Access &local&12&12&3&6\\
							&outside&1&1&1&0\\\hline\hline
							{\bf  (B)} &&	\multicolumn{2}{c}{FBO} & \multicolumn{2}{c}{NFBO}\\
							Attribute	&Level&part-worth MI & PAPRIKA MI*	& part-worth MI &  PAPRIKA MI\\\hline
							& $\ge 75\%$ &7 & 8 & 5 & 5 \\
							 Funding&$\sim 50\%$ & 2 & 1& 0 & 1 \\
							&$<25 \%$ & 4 & 3 & 1 & 0 \\\hline
							& IASC 3 & 4 &	9 &	6 &	3 \\
							Disaster Response Type&IASC 2&6&0 &0 & 2\\
							& IASC 1 or undeclared &3 & 3 & 0 & 1 \\\hline
							& clear & 12 &9.5**&6&6\\
							Assessed Need& optional&1&2.5**&0&0\\
							& unknown&0&0&0&0\\\hline
							&none&0&2&0&1\\
							Community Access &local&12&8&6&4\\
							&outside&1&2&0&1\\\hline\hline
							{\bf (C)} &&	\multicolumn{2}{c}{FBO} & \multicolumn{2}{c}{NFBO}\\
							Attribute	&Level&sample MT & population MT***	& sample MT&  population MT***\\\hline
							& $\ge 75\%$& 0 & .04$\pm$.07& .17 & .17$\pm$.09 \\
						Funding & $\sim 50\%$ & .46 & .45$\pm$.07& .33 & .33$\pm$.09 \\
							&$<25 \%$ & .54& .51$\pm .07$ & .5 & .5 $\pm .09$ \\\hline
							& IASC 3 & 0 &	.04$\pm$.06 &	.5 &	.5$\pm$ .09 \\
							Disaster Response Type& IASC 2&.85&.80$\pm$.06 &.33 & .33$\pm$.09\\
							& IASC 1 or undeclared &.15 & .16 $\pm$.06 & .17 & .17$\pm$.09 \\\hline
							& clear & .46 &.45$\pm$.07&.83&.75$\pm$.08\\
						Assessed Need & optional&.54&.51$\pm$.07&.17&.17$\pm .08$\\
							& unknown&0&.04$\pm$.07&0&.08$\pm$.08\\\hline
							&none&0&.04$\pm$.05&.33&.33$\pm.09$\\
							Community Access&local&.92&.86$\pm$.05&.5&.5$\pm$.09\\
							&outside&.08&.10$\pm$.05&.17&.17$\pm$.09\\\hline\hline
						\end{tabular}
						\flushleft 
						{\tiny	 $^{*}$ One respondent with inconsistent choice task data and hence empty feasible ranking set was removed from the sample.}\\
						{\tiny	 $^{**}$ Special cases may arise in determining a respondent's highest ranked levels and constructing the sample frequency distribution:}
						
						\begin{itemize}
							\item if two levels in the same attribute are ranked highest the same number of times in a respondent's feasible ranking set, then each level is counted .5 in the sample frequency distribution;
							\item if all three levels are ranked highest the same number of times, then they are counted .33 each in the sample frequency distribution;
							\item the above cases may necessitate the rounding of the final counts in the sample frequency distribution to the nearest whole number. If the counts of two levels end in .5, then we split the analysis into two cases, rounding one level up, and the other down; and
							\item if the respondent's feasible ranking set is empty, which indicates an inconsistency in their response, the respondent is removed from the sample (our focus and assumption is expert decision-making), and the sample size $n$ is reduced by 1. This occurred once in our disaster relief survey.
						\end{itemize}
						{\tiny	 $^{***}$ Error measures are explained in the Appendix.}
					\end{table}
	
			 Section (B) of Table \ref{MTMI}  compares the MI levels obtained by part-worth utilities with the MI levels obtained using the PAPRIKA method. Note that PAPRIKA confirms the prevalence of a MI funding level of at least 75\% for both FBO's and NFBOs. The most observable discrepancies between part-worths and PAPRIKA occur for the MI  Disaster Response Type and Community Access. Such  discrepancies signal a need for further investigation.
		
		Section (C) of Table \ref{MTMI}  compares the sample MT level proportions with their Bayesian estimated population proportions. Note that zero counts occurred in 5 out of 8 sample frequency distributions of MT levels. Using a Bayesian rather than Maximum Likelihood estimator, these levels which were zero counts in the sample are estimated to have a positive proportion in the population (see Appendix). Specifically, it is estimated that 4\% of FBOs have at least 75\% in hand at the time of the Go/No-Go decision, 4\% of FBOs most typically respond to the largest scale (IASC 3) disasters, 4\% of FBOs and 8\% of NFBOs do not know if their services are needed, and 4\% of FBOs most typically do not have an established access to the affected community.

	\section{Sensitivity Analysis}

To check the effect of a non-standard use of the Build-Your-Own  (BYO) question to determine most typical (MT) rather than most ideal (MI) levels, we performed a Monte Carlo simulation of a generic survey with attributes $A$,$B$,$C$,$D$ each of whose levels are indicated with the subscripts 1,2 and 3. We assumed without loss of generality that the respondent's partworth utility for each level decreased as the level index value increased. Hence $A_1,B_1,C_1$ and $D_1$ were the MI levels.  In cases where the total utility of 2 paired profiles were equal, we assumed the attributes were ranked in order of increasing importance, $A$, $B$, $C$ and $D$,  so the profile with higher utility level for attribute $A$ was chosen; it the $A$ levels were the same, the profile with higher utility level for the $B$ attribute was chosen, and so on.  Table \ref{MC} indicates that  for a tournament whose profiles are based on the standard use of the BYO to determine the MI levels, the PAPRIKA method correctly estimates each MI level with probability approximately .72.
Using a pre-specified or randomized  non-standard BYO  to determine the MT levels rather than MI levels, the probability reduced slightly to approximately .68.  Note that the probability that PAPRIKA correctly determines an expert respondent's MI levels depends on the structure of the survey and choice task tournament, but not on the sample size $n$.  PAPRIKA is therefore particularly suitable as a validation tool for  MI levels obtained by part-worths for small samples.

\begin{table}[t]
	\scriptsize
	\captionsetup{width=.9\linewidth}
	\caption{\\ Probability that PAPRIKA correctly predicts  MI levels $A_1,B_1,C_1$,and $D_1$. For each choice of BYO, 10,000 tournaments were generated with tournament profiles obtained by randomly varying two of the BYO levels.   \\}
	\centering
	\begin{tabular}{ccccccccccccc}
	
		& \multicolumn{4}{c}{\textbf{(i) BYO = $A_1B_1C_1D_1$}} & \multicolumn{4}{c}{\textbf{(ii) BYO=$A_2B_2C_2D_2$}} & \multicolumn{4}{c}{\textbf{(iii) Random BYO }} \\
		& $A_1$ &$B_1$ & $C_1$ & $D_1$ &$A_1$ &$B_1$ & $C_1$ & $D_1$ &  $A_1$ &$B_1$ & $C_1$ & $D_1$ \\
		\hline
		probability & .714 & .719 & .712 & .724 &.680 & .687 & .677 & .687 & .687 & .680 & .679 & .674 \\
			\hline
	\end{tabular}
	\label{MC}
\end{table}

\section{Discussion}

This Note introduces a simple method which uses an ACBC survey with a small sample size $n$ from a small population $N$ to

\begin{itemize}
\item compare  respondents' most typical (MT) levels  (obtained by non-standard use of the Build Your Own (BYO) question) with their most ideal (MI) levels (obtained  by means of part-worth utilities);
\item validate each respondent's MI levels determined by part-worths using the PAPRIKA method; and
\item explore differences in MT and MI levels for two related populations.
\end{itemize}
             
 Table   \ref{summary1} gives a quick overview of our method's goal, procedure for implementation, and benefits.           
             	\begin{table}[t]
             	\scriptsize
             	\captionsetup{width=.95\linewidth}
             	\caption{\\Overview of MT/MI Estimation Method for an ACBC survey of a small population $N$ and sample size $n$.\\}
             	\centering
             	\label{summary1}
             	\begin{tabular}{l}
             		GOAL\\
             		Accurate estimation of most typical (MT) and most ideal (MI) levels for each expert decision-maker respondent in a small sample $n$ from a small population $N$.\\\hline
             		PROCEDURE\\
             		1. Determine MT levels using the Build-Your-Own question.\\
             		2. Determine MI levels using choice-task data to estimate part-worth utilities.\\
             		3. Use PAPRIKA to validate MI levels for each respondent.\\
             		4. Compile sample frequency distributions of MT/MI levels.\\
             		5. Use a Bayesian method to estimate population frequency distributions of MT/MI levels. (Recommended for estimation of zero counts.) \\\hline
             		BENEFITS\\
             		1. PAPRIKA validation of a respondent's  MI levels is suited for small sample size $n$.\\
             		2. Lack of PAPRIKA validation of MI levels suggests a need for further investigation of attribute.\\
             		3. MT/MI sample frequency distributions allow for both within-population and between-population comparisons.

             	\end{tabular}
             \end{table}

PAPRIKA validation of the MI levels is particularly appropriate if each respondent is an expert decision-maker whose choice task data is non self-contradicting.  For important small populations such as CEOs of large companies or mayors of large U.S. cities, the expert decision maker assumption is reasonable. In the case of our survey, out of 19 respondents, 16 had over 15 years experience managing disaster relief.  This was borne out by the fact that only 1 of 19 choice task data sets was inconsistent.  In any case, binary choice tasks and a single elimination tournament structure, or a survey using  BestWorst Scaling (Marley and Islam 2012) can reduce the possibility of self-contradictory choice task data. Wherever there is obvious disagreement between PAPRIKA and part-worth MI levels of an attribute, a need for further investigation of that attribute is advisable.  For example, our survey suggests a need to investigate further  the most ideal disaster response type for both FBOs and NFBOs, as well as how prevalent Community Access through local partners is deemed most ideal by both groups.

A survey of a small population $N$ with a small sample size $n$ may well include zero count MT/MI levels. This was the case for 9 out of the 16 sample frequency distributions of MT/MI levels for FBOs and NFBOs. Whereas maximum likelihood estimation of a hypergeometric distribution will always estimate persistence of zero counts for the entire population, a Bayesian approach may assign a small but positive proportion of the population to a zero count level in a sample (see the Appendix).

 In view of a bias towards the BYO in randomly varying 2 of the 4 BYO attribute levels to generate choice task profiles (Louviere 2008), a Monte Carlo sensitivity analysis indicated that for our survey structure (4 attributes each with 3 levels), using round of 16 single-elimination choice task data for a small sample size $n$ and non-standard choice of BYO, PAPRIKA  correctly determines each of a respondent's four MI levels with probability approximately .7. For  broader applicability of our approach to determining MT/MI levels, important questions should be investigated. Assuming that each respondent is an expert decision maker whose choice-task data is non-contradictory,  given an ACBC survey structure with attribute $A_i$ having $a_i$ levels ($i=1,...,k$), are there  optimal choice task designs for obtaining MI levels by both part-worth utilities and PAPRIKA? For such optimal designs, what are the probabilities that part-worth utilities and PAPRIKA estimate MI levels correctly? How are optimal designs affected  under relaxed assumptions on $n$, $N$, and expert decision-making respondents?

While the larger population of $N=49$ (Christian) international FBOs was obtained by an extensive search, the smaller population of $N=12$ NFBOs was obtained by a simple membership criteria in National Voluntary Organizations Active in Disaster.  Table \ref{summary} gives a summary comparison of these two small populations. Similarity and differences between FBO and NFBO experience (MT levels) and aspirations (MI levels) suggest both opportunities and barriers for fuller engagement of disaster relief organizations in the most difficult disaster responses. For example,  we estimate that 30\% of FBOs would ideally like to participate in the largest type of response (IASC 3)  though this was not the MT level for any of the sample FBOs. Moreover, half the FBOs are estimated to respond with ideally  50\% or less funding in hand.  However, only 7\% of FBOs are estimated to typically or ideally access an affected community through an outside organization, and an even smaller percentage of FBOs is estimated to ideally/typically respond when there is no pre-existing partnership.  To facilitate greater FBO response to IASC 3 disasters in regions which are unfamiliar to them,  prior development of collaborative relationships with   major NFBOs   may be difficult, but potentially more strategic preparation than increasing FBO funding for such responses.

 \begin{table}[t]
 	\captionsetup{width=.7\linewidth}
 	\scriptsize
 	\caption{{\small \\ Summary Comparison of FBO and NFBO MT and MI levels. }\hspace{1in}.\\}
 	\centering
 	\label{summary}
 	
 	\begin{tabular}{lll}
 		Attribute	& Commonality	& Differences\\\hline
 		{\bf Funding}& \multicolumn{1}{m{1.5in}}{Both FBOs and NFBOs typically decide to respond with only up to half of the funding needed in hand.}  & \multicolumn{1}{m{1.5in}}{While roughly half of the FBOs do not ideally need the highest funding level, a large majority of the NFBOs view having at least 75 \% as their ideal.}  \\\\
 		{\bf Disaster Response Type}& \multicolumn{1}{m{1.5in}}{  Similar to all NFBOs, roughly a third of the FBOs would ideally like to be involved in the largest category of response (IASC 3).}   &\multicolumn{1}{m{1.5in}}{  Whereas FBOs typically respond only to medium- and small-scale disasters, half the NFBOs typically respond to the largest scale disasters.}   \\\\
 		{\bf Assessed Need}& \multicolumn{1}{m{1.5in}}{ FBOs and NFBOs both consider clear Need assessment  as most ideal.} & \multicolumn{1}{m{1.5in}}{ Whereas  NFBOs typically respond only when there is a clear Need assessment, roughly half the FBOs typically respond when their services are optional.}\\\\
 		{\bf Community Access:} & \multicolumn{1}{m{1.5in}}{ Both FBOs and NFBOs view  local partner Access as most ideal.}  & \multicolumn{1}{m{1.5in}}{ Whereas FBOs typically respond only when they have local partners, half the NFBOs typically respond  without a local partner.} \\\hline
 	\end{tabular}
 \end{table}

\subsection*{Acknowledgements}

Special thanks to Dr. Erica Gralla, Dr. Michael Veatch, Dr. Jarrod Goentzel, Dr. Darcie Delzell, Mr. David Husby, Mr. and Mrs. Levi and Diane Velasco, Mr. David Bakalemwa, Ms. Claire Browning, Ms. Ming-Hsuan Chuang,
Mr. Daniel Daum, Ms. Michaela Flitsch, Ms. Kelli Geney, Ms. Zoe Kallus, Mr. Timotius Kartajiwaya, Mr. Jonathan Larson, Ms. Courtney Linscott, Ms. Sara Magnuson, Mr. Mark Nussbaum,   Mr. Zach Oslund, Mr. Matthew Rueger, Mr. Nate Schatz,  Mr. Hudson Thomas, Mr. Nick Varberg, and Ms. Joyce Yan for their assistance with the design, implementation, and data processing of the disaster relief survey discussed in this paper.

	\newpage
	\section*{Appendix: Zero Count Estimation}
	In light of the fact that zero counts arose in 9 of 16 sample frequency distributions of MT/MI levels for the FBOs and NFBOs, we explain why we used a Bayesian rather than maximum likelihood approach to zero count estimation of the population MT/MI level distributions (see Table \ref{MTMI}).  Using a weighted mean absolute error to measure estimation error, maximum likelihood estimation (Method 1) is not the best choice of estimator, even among estimators which maintain the zero count level for the entire population.  The Bayesian estimator we have employed (Method 2)  is one which minimizes the weighted mean absolute error and allows a zero count MT/MI level occurring in a small sample to have a positive probability of occurring in the population.  We provide mathematical details about each method below.
		
	\subsubsection*{Method 1: Maximum Likelihood Estimation (MLE)}
For an ACBC survey attribute which has $m$ levels, a sample frequency distribution of MT (or MI) levels has the form $S=\langle n_1,...,n_m\rangle$, where $n_i$ is the number of respondents whose MT (or MI) level is level $i$. (Levels are numbered 1,2,...,$m$ but may be categorical rather than ordinal.) For a  population with known size $N$,  the actual population frequency distribution of the MT (or MI) level under consideration is  $P=\langle N_1,...,N_m\rangle$, where  we assume no knowledge about $P$ other than $N_i\ge n_i$ ($i=1,...,m$)  and  $\sum_{i=1}^m N_i=N$.  The maximum likelihood estimator (MLE) has the form $\hat{P}^*=\langle \hat{N}_1^*,..., \hat{N}_m^*\rangle$  where $\sum_{k=1}^m N_i^*=N$ and $N_i^*\ge N_i$ for $i=1,...,m$. For a known population size $N$, the MLE can be obtained by means of a probability factor table (Oberhofer and Kaufman 1987) whether or not $S$ contains a zero count (i.e. $n_i=0$ for some $i\in\{1,...,m\}$).
	
	If a sample of size $n$ is a random draw from a population of size $N$, the observed numbers $n_i$ ($i=1,...,m$) are joint multivariate hypergeometrically distributed. 	While the theory of multivariate hypergeometric distributions is well-developed for large sample size $n$ and asymptotic behavior as both $n$ and $N\rightarrow\infty$ (Janardan 1976, Hartley and Rao 1968),  our focus is on zero counts arising for small  $n$ and a known, small fixed value for $N$. For the latter, we use a probability factor table 
	(Table \ref{PFT}) to construct an MLE population frequency distribution estimate $\hat{P}^*=\langle \hat{N}_1^*,...,\hat{N}_m^*\rangle$. 
	
		\begin{table}[t]
		\captionsetup{width=.8\linewidth}
		\caption{\\A probability factor table can be used to construct the maximum likelihood estimation (MLE) for a multivariate hypergeometric distribution with a known population size $N$.}
		\begin{center}
			\label{PFT}
			\scriptsize
			\begin{tabular}{ccccccc}\hline
				actual number in population&$N_1$&$N_2$&...&$N_i$&...&$N_m$\\\hline
				estimated number in population&$\hat{N}_1=n_1+x_1$&$\hat{N}_2=n_2+x_2$&...&$\hat{N}_i=n_i+x_i$&...&$\hat{N}_m=n_m+x_m$\\\hline
				$\vdots$&$\vdots$&$\vdots$&...&$\vdots$&...&...\\
				+$j$&$f_{1j}=1+\frac{n_1}{j}$&$f_{2j}=1+\frac{n_2}{j}$&...&$f_{ij}=1+\frac{n_i}{j}$&...&$f_{mj}=1+\frac{n_m}{j}$\\
				$\vdots$&$\vdots$&$\vdots$&...&...&...&$\vdots$\\
				+2&$f_{12}=1+\frac{n_1}{2}$&$f_{22}=1+\frac{n_2}{2}$&...&$f_{i2}=1+\frac{n_i}{2}$&...&$f_{m2}=1+\frac{n_m}{2}$\\
				+1&$f_{11}=1+\frac{n_1}{1}$&$f_{21}=1+\frac{n_2}{1}$&...&$f_{i1}=1+\frac{n_i}{1}$&...&$f_{m1}=1+\frac{n_m}{1}$\\
				+0&$f_{10}=1$&$f_{20}=1$&...&$f_{i0}=1$&...&$f_{m0}=1$\\\hline
				number in sample &$n_1$&$n_2$&...&$n_i$&...&$n_m$\\\hline
				most preferred level&$A_1$&$A_2$&...&$A_i$&... &$A_m$\\
			\end{tabular}
		\end{center}
	\end{table}   

	For each $i$ between 1 and $m$, let $x_i$ be the difference between the predicted number $\hat{N}_i$ in the population who most prefer level $A_i$ and the number $n_i$ observed in the sample with that same level preference.  Note that

			\begin{eqnarray*}
				C(n_i+j,n_i) & = & \frac{(n_i+j)!}{n_i!j!}\\
				&= &\frac{(n_i+j)(n_i+j-1)...(n_i+1)}{j(j-1)...1}\\
				&=&(1+\frac{n_i}{j})(1+\frac{n_i}{j-1})...(1+\frac{n_i}{1})\\
				&=&\Pi_{j=0}^m f_{ij},
			\end{eqnarray*}
			
			{\flushleft where} $f_{i0}=1$ and  $f_{ij}=1+\frac{n_i}{j}$.  The probability of a population $\hat{P}$ with $\hat{N}_i=n_i+x_i$ having dominant attribute level $A_i$ and giving rise to the observed sample value $n_i$ ($i=1,...,m$) is equal to
			\begin{equation}
			\frac{\Pi_{i=1}^m(\Pi_{j=0}^{x_i} f_{ij})}{C(N,n)},
			\end{equation}
			{\flushleft where} $f_{ij}=1+\frac{n_i}{j}$, and the denominator $C(N,n)$ does not depend on the choice of $x_i$.  An MLE distribution is therefore obtained by choosing a set $X^*$ which contains $N-n$ largest factors $f_{ij}$ appearing in Table \ref{PFT}.  The choice of factors is not always unique,  but is simplified by the fact that the factors in each column are decreasing as one moves up the column (i.e. $f_{ij}=1+\frac{n_i}{j}$ is a decreasing function of $j$). Let ${x}^*_i$ denote the number of largest factors in $X^*$ selected from column $i$.  Then an MLE estimator $\hat{P}^*$  for the population distribution $P$ is  given by
			\begin{equation}
			\hat{P}^*=\langle \hat{N}_1^*,...,\hat{N}_m^*\rangle=\langle n_1+x_1^*,...,n_m+x_m^*\rangle.
			\end{equation}
			
			One problem with the use of an MLE estimator $\hat{P}^*$ in a small population/small sample survey is that $n_i$ might equal 0 for some $i$.  (In our ACBC survey of disaster relief organizations, such zero counts occurred in 9 out of 16 sample frequency distributions considered.) If $n_i=0$, the  corresponding MLE estimate is $\hat{N}_i^*=0$. This must be the case since $f_{ij}=1+\frac{n_i}{j}=1$ for all $j$ if $n_i=0$, and $f_{ij}>1$ if $n_i>0$.  A simple example suggests why this is problematic.  Suppose attribute $A$ has $m\ge 3$ levels. If a sample has size $n=2$, at least $m-2$ levels of $A$ must be  0-levels.  MLE says these levels are also 0-levels for the entire population  for every value of $N>2$.     MLE estimates the dominant levels of just two people will be held by the entire population, no matter how large the population $N$ and how many levels $m\ge 3$ are considered.
			
	Let us illustrate how to use a probability factor table to obtain the MLE estimate $\hat{P}^*=\langle \hat{N}_1^*,..., \hat{N}_m^*\rangle$ for the attribute $\mathcal{A}=$External Funding.  Three funding levels ($m=3$) were considered: $\ge 75\%$ ($A_1$); \hspace{.05in} $\sim 50\%$ ($A_2$); \hspace{.05in} $<25\%$ ($A_3$). For the NFBOs surveyed, the population size was $N=12$ and sample size $n=6$.  The number $n_i$ of sample respondents with funding BYO level $A_i$ was $n_1=1,n_2=2,$ and $n_3=3$.
	
	Table \ref{NGOBYOFunding} shows the corresponding probability factor table. Observing that since the  6 largest factors are $f_{11}=2, f_{21}=3,f_{22}=2, f_{31}=4,f_{32}=2.5, f_{33}=3,$ we use $x_1^*=1$ (one factor chosen from column 1), $x_2^*=2$ (two factors chosen from column 2), and $x_3^*=3$ (three factors chosen from column 3). The resulting MLE is
	$\hat{N}_1^*=n_1+x_1^*=1+1=2$, $\hat{N}_2^*=n_2+x_2^*=2+2=4$, and $\hat{N}_3^*=n_3+x_3^*=3+3=6$.
	
		\begin{table}[t]
		\captionsetup{width=.9\linewidth}
		\centering
		\caption{\\A probability factor table can be used to construct the Non-Faith Based Relief Organization (NFBO) most typical (MT) funding maximum likelihood estimation (MLE) for a multivariate hypergeometric distribution with known population size $N=N_1+N_2+N_3$.}
		\label{NGOBYOFunding}
		\scriptsize
		\begin{tabular}{cccc}\hline
			actual number in population&$N_1$&$N_2$&$N_3$\\\hline
			MLE predicted number&$\hat{N}_1^*=n_1+x_1=1+1=2$&$\hat{N}_2^*=n_2+x_2=2+2=4$&$\hat{N}_3^*=n_3+x_3=3+3=6$\\\hline
			&&&\\
			+3&$f_{13}=1+\frac{1}{3}=4/3$&$f_{23}=1+\frac{2}{3}=5/3$&{\bf $f_{33}=1+\frac{3}{3}=2$}\\
			+2&$f_{12}=1+\frac{1}{2}=1.5$&{\bf $f_{22}=1+\frac{2}{2}=2$}&{\bf $f_{32}=1+\frac{3}{2}=2.5$}\\
			+1&{\bf $f_{11}=1+\frac{1}{1}=2$}&{\bf $f_{21}=1+\frac{2}{1}=3$}&{\bf $f_{31}=1+\frac{3}{1}=4$}\\
			+0&$f_{10}=1$&$f_{20}=1$&$f_{30}=1$\\\hline
			number in sample &$n_1=1$&$n_2=2$&$n_3=3$\\\hline
			BYO level&$A_1$&$A_2$&$A_3$\\
		\end{tabular}
	\end{table}

	\subsubsection*{Method 2: Weighted Mean Absolute Error (WMAE) Minimization}

	To address the inability of MLE to adequately estimate zero counts, we introduce the following Bayesian alternative,  one which is easily implementable on a computer. 	As before, let $P=\langle N_1, N_2,...,N_m\rangle$ be any admissible population frequency distribution and $\hat{P}=\langle\hat{N}_1, \hat{N}_2 ,...,\hat{N}_m\rangle$ any admissible estimate of the population frequency distribution.  By admissible we mean that the vector components sum to $N$ and each $i^{th}$ component is greater than or equal to $n_i$. We then define the mean absolute error $E(P,\hat{P})$ as
	\begin{equation}
	E(P,\hat{P}) = \frac{1}{m}\sum_{i=1}^m |N_i - \hat{N}_i|.
	\label{E1}
	\end{equation}
	This error is a simple average of how far off the estimate $\hat{N}_i$ is from the actual $N_i$. Even without knowing the actual values for the $N_i$, we may obtain a simple bound on $E(P,\hat{P})$:
	\begin{equation}
	E(P,\hat{P}) = \frac{1}{m}\sum_{i=1}^m |N_i - (n_i+x_i)| \le \frac{1}{m} [\sum_{i=1}^m (N_i-n_i) + \sum_{i=1}^m x_i ] = \frac{2}{m} (N-n).
	\label{E1}
	\end{equation}
	
	More to the point, without knowing the actual values of $N_i$, we can use a weighted mean absolute error (WMAE) to determine an optimal population estimate, especially in cases where there are zero counts. Let $p_k$ be the probability that a sample distribution $S=\langle n_1,...,n_m\rangle$ arises from an admissible population $P_k$, and let $p=\sum_{\textup{admissible P}_k} p_k$.
	Define the weighted mean absolute error $\bar{E}(\hat{P})$ as
	\begin{equation}
	\bar{E}(\hat{P}) =\sum_{\textup{admissible} P_k} \frac{p_k}{p} E(P_k,\hat{P}).
	\end{equation}
	
	The WMAE $\bar{E}(\hat{P})$ averages the error $E(P_k,\hat{P})$ over all admissible populations $P_k$, with the weight determined by the probability that the population $P_k$ gives rise to the observed sample.
	A table or graph of   $\bar{E}(\hat{P})$ can then be used to show the range of WMAE's that can occur and also identify the  approximation $\hat{P}^\dagger $ for $P$ which minimizes the WMAE.  $\bar{E}(\hat{P}^\dagger )$  then serves as a much better estimate than (\ref{E1}) for how far off each level estimate $\hat{N}_i^\dagger$ is from the actual value for $N_i$.
	
		An estimator which minimizes the weighted mean absolute error  is a type of minimax estimator (Jokiel-Rokita 1998), a Bayesian approach  which assigns probability  weights to possible population frequency distributions and seeks an estimator which minimizes the ``risk'', meaning the maximum expected estimation error, where the estimation error (``loss'') is given by a quadratic function.   For our context of small $n$ and $N$, for simplicity we chose to use an absolute (rather than quadratic) error measure, and also locate the weighted mean absolute error minimizing estimator using a graphical method rather than by derivation of a mathematical formula (Fink 1997).

\begin{figure}[!htpb]
		\centering
		\includegraphics[width=5.75in, height=2.in]{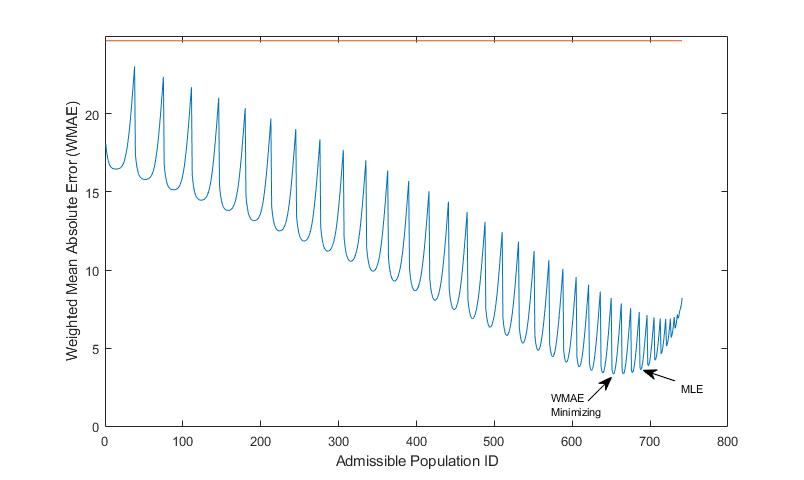}
		\caption{{\small Plot of the WMAE for each admissible estimator $\hat{P}$ of the FBO population ideal scale assessment level with sample values $n_1=9$, $n_2=0$, and $n_3=3$, The  population  $\hat{P}^\dagger=\langle \hat{N}_1^*=34$, $\hat{N}_2^*=2$,  $\hat{N}_3^*=13\rangle$  (ID \#653) has the minimum WMAE $\bar{E}(P^\dagger)=3.35$.
				The MLE estimator (ID \#687) has a larger WMAE $\bar{E}(P^*)=3.73$.  By (\ref{E1}), the WMAE in all cases must be bounded by  $\frac{2}{3}(N-n)=24.67$} (horizontal line).}
		\label{MLError}
	\end{figure}

	The following example shows the difference between the MLE estimator $\hat{P}^*$ and WMAE minimizing estimator $\hat{P}^\dagger$ in the case where a zero count occurs. For the  attribute $\mathcal{A}=$Disaster Scale, three levels were considered: IASC 3 ($A_1$); IASC 2 ($A_2$); IASC 1 or undeclared  ($A_3$).  In this case, our FBO population size was $N=49$ and our sample size was reduced from $n=13$ to $n=12$ due to an empty FR set (see Section 2.3). The MI sample frequency distribution was in this case $n_1=9$, $n_2=0$, and $n_3=3$. Figure \ref{MLError} shows the WMAE $E(\hat{P})$ for all admissible  FBO population estimators $\hat{P}=\langle \hat{N}_1,\hat{N}_2,\hat{N}_3 \rangle$. The estimate $\hat{P}^\dagger=\langle \hat{N}_1^{\dagger}=34, \hat{N}_2^{\dagger}=2, \hat{N}_3^{\dagger}=13 \rangle$ (ID \#653) had the minimum WMAE of $\bar{E}^{\dagger}=3.35$.  The MLE estimate $\hat{P}^*=\langle \hat{N}_1^*=37,\hat{N}_2^*=0,\hat{N}_3^*=12\rangle$ has a higher WMAE of  $\bar{E}^*=3.73$.  Note  that the inequality (\ref{E1}) only guarantees that $E(\hat{P})\le \frac{2}{m}(N-n)=\frac{2}{3}(49-12)=24.67$ for all admissible populations $\hat{P}$, so $\bar{E}^{\dagger}=3.35$ provides a much better error estimate.

	Zero counts have been investigated in a wide variety of fields such as education (number of courses students failed, Salehi and Roudbari 2015), health (decayed or missing teeth, Moghimbeigi et al. 2008), linguistics (frequency of word usage, Brysbaert and Diependale 2013), and criminology (counts of arson in a police precinct, Zhu 2012). Such problems involve different distributions (eg. Poisson and negative binomial distributions) and various heuristic strategies such as elimination of the zero count category or merging it with others, adding one to the count in each category, or
	considering the sample as including a subgroup with zero probability of belonging to the zero level category (Brysbaert and Diependale 2013, Allison 2012).

	\section*{References}
	
	\small
	
	  Allison, P. 2012. Do we really need zero-inflated models? https://statisticalhorizons.com/zero-inflated-models Accessed 5/18, 2019.
	 
 Brysbaert, M., and  Diependaele, K. 2013. Dealing with zero word frequencies: A review of the existing rules of thumb and a suggestion for an evidence-based choice. \emph{Behavior Research Methods} 45(2):422-430.

		CCCU n.d. Networking Grants. Available from https://www.cccu.org/facultygrants/networking-grants/\#heading-past-grant-recipients-5  Accessed 12/14/2019.

	Cunningham, C., Deal, K.,  Chen, Y. 2010. Adaptive Choice-Based Conjoint Analysis: A New Patient-Centered Approach to Assessment of Health Service Preferences. \emph{The Patient}, Springer, New York. 3(4):257-273.

	Devers,K., Gray, B., Ramos, C., Shah, A., Blavin, F., Waidmann, T.  2014. The Feasibility of Using Electronic Health Data for Research on Small Populations, Available from https://www.urban.org/research/publication/feasibility-using-electronic-health-data-research-small-populations
	
	Douglas, C. 2017. Your mayor makes how much? Here's the breakdown, and the numbers might surprise. Available at
	https://www.bizjournals.com/bizjournals/news/2017/01/17/your-mayor-makes-how-much-heres-the-breakdown-and.html
	
		Equilar 2019. Top 25 Survey. Available at https://www.equilar.com/survey.html

	Fink, D. 1997. A Compendium of Conjugate Priors. Available at https://www.johndcook.com//CompendiumOfConjugatePriors.pdf.

	Gingerich, T., Moore, D., Brodrick, R., Beriont, C. 2017. Local Humanitarian Leadership and Religious Literacy: Engaging with Religion, Faith, and Faith Actors. Harvard Divinity School report http://rlp.hds.harvard.edu/files/hds-rlp/files/rr-local-humanitarian-leadership-religious-literacy-310317-en.pdf

	Gralla, E., Goentzel, J., Fine G. 2014. Assessing trade-offs among multiple objectives for humanitarian aid delivery using expert preferences.
	\emph{Production and Operations Management}, Springer-Verlag Berlin 23(6), 978-989.
	
	Hammond, P.. Zank, H. Rationality and Dynamic Consistency Under Risk and Uncertainty. in \emph{Handbook of the Economics of Risk and Uncertainty}, Vol 1, 2014, 41-47.
	
	Hansen, P. and Ombler, H. 2009. A new method for scoring additive multi-attribute value models using pairwise rankings of alternatives. Available at
	 https://doi.org/10.1002/mcda.428

	Hartley, H.O. and Rao, J.N.K. 1968. A new estimation theory for sample surveys. Biometrika, 55(3), 547-557.
	
	Henning, J. 2014. The paradox of surveys of small population sizes. Available at http://researchaccess.com/2014/10/surveys-of-small-population/.
	
	Janardan, K. 1976. Certain Estimation Problems for Multivariate Hypergeometric Models. Ann. Inst. Statist. Math. 28(Part A), 429-444.
	
	Johnson, R. 1975. A Simple Method of Pairwise Monotone Regression. \emph{Psychometrika}, 40(2), 163-168.
	
	Jokiel-Rokita, A. 1998. Minimax Prediction for the Multinomial and Multivariate Hypergeometric Distributions. Applicationes Mathematicae, 25(3), 271-283.
	
	Kirkendall, J., White, J., Rapporteurs. 2018. Improving Health Research of Small Populations: Proceedings of a Workshop. Available at https://www.nap. edu/read/25112/chapter/1.
	
	Lighthouse Studio Help, Analysis with Tiny Sample Sizes. Available at http://www.sawtoothsoftware.com/help/lighthouse-studio/manual/index.html?analysiswithtinysamplesize.html.

	Louviere, J., Street, D., Burgess, L., Wasil, N., Islam, T., Marley, A. 2008.
	Modeling the choices of individual decion-makers by combining efficient choice experiment designs with extra preference information. \emph{Journal of Choice Modeling}, 1(1), 128-164.
	
	Marley, A.,  Islam T., 2012. Conceptual Relations Between Expanded Rank Data and Models of the Unexpanded Rank Data. \emph{Journal of Choice Modelling}, 5(1),  38-80.
	
	Marshall, P., Bradlow, E. 2002. A Unified Approach to Conjoint Analysis Models. \emph{Journal of the American Statistical Association}, 97(459), 674-682.
	
	Maydeu-Olivares, A. and B\"{o}ckemholt, U. 2005. Structural Equation Modeling of Paired-Comparison and Ranking Data. \emph{Psychological Methods}, 10(3), 285-304.
	
		Means, B., Toyama, Y., Murphy, R., Bakia, M., Jones, K. 2010. Evaluation of Evidence-Based Practices in Online Learning: A Meta-Analysis and Review of Online Lwearning Studies. https://www2.ed.gov/rschstat/eval/tech/evidence-based-practices/finalreport.pdf Accessed 5/15/2019.

	Moghimbeigi, A., Eshraghian, M., Mohammad, K., Mcardle, B. 2008. Multilevel Zero-Inflated Negative Binomial Regression Modeling for over-Dispersed Count Data with Extra Zeros. \emph{Journal of Applied Statistics}, 35(10),1193–1202.

	Oberhofer, W., Kaufman, H. 1987.  Maximum Likelihood Estimation of a Multivariate Hypergeometric Distribution. \emph{Sankhya: The Indian Journal of Statistics, Series B (1960-2002)}, Indian Statistical Institute, 49(2), 188-191.
	
	Orme, B.K. 2014. \emph{Getting Started with Conjoint Analysis.} Sawtooth Software, Inc.
	
	Orme, B.K., Chrzan, K. 2017. \emph{Becoming an Expert in Conjoint Analysis: Choice Modeling for Pros.} Sawtooth Software, Inc.
	
	Orme, B.K. and Johnson, R. M. 2008. Testing Adaptive CBC: Shorter Questionnaires and BYO vs. ``Most Likelies''.
		 https://www.sawtoothsoftware.com/support/technical-papers/adaptive-cbc-papers/testing-adaptive-cbc-shorter-questionnaires-and-byo-vs-most-likelies-2008  Accessed 5/10/2019.
	
	Rao, V. R. 2014. \emph{Applied Conjoint Analysis}. Springer.
	
	Rossi, P., Allenby, G., McCulloch R. 2005. \emph{Baysian Statistics and Marketing.} John Wiley \& Sons, Ltd.
	
		Russell Reynolds Associates, 2010. In Touch with the Board.  https://www.russellreynolds.com/sites/default/files/Board\_Evaluation.pdf  Accessed 6/14/2019.
		
			Salehi, M., Roudbari, M. 2015. Zero inflated Poisson and negative binomial regression models: application in education. \emph{Medical Journal of the Islamic Republic of Iran}, vol. 29, 1-6.
		
	Sawtooth Software,  2009.  CBC/HB v5 5.0 https://www.sawtoothsoftware.com/download/ssiweb/CBCHB\_Manual.pdf  Accessed 5/13/2019.
	
	Sawtooth Software,  2017. Lighthouse Studio v9.5: Software for Web Interviewing and Conjoint Analysis. Sawtooth Software, Inc.

	Sawtooth Software, 2013. The MaxDiff System Technical Paper. https://www. sawtoothsoftware.com/download/techpap/maxdifftech.pdf Accessed 11/27/2018.

	Thurstone, L. L. 1927. A Law of Comparative Judgment, \emph{Psychological Review}, 4, 273-286.
	
	Veatch. M. and Reimel H. 2017. Informed Compassion: the Interplay of Faith Perspectives and Humanitarian Logistics.https://drive.google.com/file/d/0B6hH5h4c9fSNS3F4ZlY3MkhaNnM/view?usp=sharing
   Accessed 12/16/2019.
	
	Zakour, M. J., Harrell, E.B. 2003. Access to disaster services: Social work interventions for vulnerable populations. Journal of Social Service Research, 30(2), 27-54.
	
	Zhu, F. 2012. Zero-inflated poisson and negative binomial integer-valued garch models. \emph{Journal of Statistical Planning and Inference}, 142(4), 826-839.

\end{document}